\documentclass[final,3p,times,twocolumn]{elsarticle}
\usepackage[utf8]{inputenc}

\usepackage{amsmath}
\usepackage{amsfonts}
\usepackage{amssymb}
\usepackage{graphicx}
\usepackage{color}
\usepackage{array,tabularx}

\def\<{\langle}
\def\>{\rangle}

\def\db{\<u^2\>}

\newcommand{\kB}{k_{\rm B}}

\begin{document}

\begin{frontmatter}
\title{Elastic models of the glass transition applied to a liquid with density anomalies}

\author{Massimo Pica Ciamarra} 
\address{Division of Physics and Applied Physics, School of
Physical and Mathematical Sciences, Nanyang Technological University, Singapore}
\address{
CNR--SPIN, Dipartimento di Scienze Fisiche,
Universit\`a di Napoli Federico II, I-80126, Napoli, Italy
}
\author{Peter Sollich}
\address{
King’s College London, Department of Mathematics, Strand, London WC2R 2LS, United Kingdom
}

\begin{abstract}
Elastic models of the glass transition relate the relaxation dynamics and the elastic
properties of structural glasses. They are based on the assumption that
the relaxation dynamics occurs through activated
events in the energy landscape whose energy scale is set by the elasticity of the material.
Here we investigate whether such elastic models describe the 
relaxation dynamics of systems of particles
interacting via a purely repulsive harmonic potential,
focusing on a volume fraction and temperature range
that is characterized by entropy--driven water--like density anomalies.
We do find clear correlations between 
relaxation time and diffusivity on the one hand, and plateau shear
modulus and Debye--Waller factor on the other,
thus supporting the validity of elastic models of the glass transition.
However, we also show that the plateau shear modulus is not related
to the features of the underlying energy landscape of the system,
at variance with recent results for power--law potentials.
This challenges the common potential energy landscape interpretation
of elastic models.
\end{abstract}


\end{frontmatter}

\maketitle

\section{Introduction}
The relaxation dynamics of supercooled liquids 
occurs through local particle rearrangements
whose energy cost is related to the
elastic properties of the material.
This suggests the existence of correlations between
the elasticity and the dynamics of supercooled liquids.
Indeed, at the local scale
particle mobilities have been found to be related to local
elastic constants~\cite{Reichman2008}.
At the macroscopic level these correlations
have stimulated the formulation
of elastic models of the glass transition,
which relate the relaxation dynamics
and the elasticity of glass formers.
For instance, according to Dyre's shoving model~\cite{Dyre06},
the relaxation occurs through local
volume-preserving events that allow the system
to transit from one potential energy minimum to a different one.
If one estimates the energy barrier separating
two energy minima within a parabolic approximation, this approach 
leads to the relation
\[
\tau \propto \exp\left(G_p V_{\rm at}/\kB T\right)
\]
between the relaxation time and the plateau shear
modulus, $G_p$, 
where $V_{\rm at}$ is an atomic volume element 
that is assumed to be temperature independent, $T$ is temperature and
$\kB$ is Boltzmann's constant. See Refs.~\cite{Puosi2012,Dyre2012,Potuzak2013}
for a discussion regarding the interpretation of $G_p$.
Recent numerical results on polymer melts~\cite{Puosi2012} and on systems
of particles interacting via inverse power law potentials~\cite{Abraham}, 
showed the possibility of connecting the plateau shear modulus
to features of the energy landscape of the system~\cite{Goldstein,DebenedettiStillinger}.
Indeed, these studies found $G_p$ to be related to the fluctuations
of the inherent shear stress, $G_p = G^{\rm IS} = \frac{V}{\kB T} \<
(\sigma^{\rm IS}_{xy})^2 \>$. Here
$\sigma^{\rm IS}_{xy}$ is the shear stress measured after quenching a system
to its inherent structure, i.e.\ the nearest potential energy minimum. Together, Dyre's shoving model and
the results of Refs.~\cite{Puosi2012,Abraham} lead to a relation between
the relaxation time on the one hand, and features of the inherent
energy landscape on the other.

A related approach to connecting the dynamics to elastic properties 
via features of the energy landscape can be motivated by an analogy
with the Lindemann melting criterion for periodic crystal
structures. Here a relaxation event is considered
to occur through local rearrangements that take place when the mean squared vibrational amplitude 
of a particle $\<u^2\>$, which is its Debye--Waller (DW) factor,
crosses some threshold $a^2$, where $a$ is of the order of the
particle size. 
If this process requires an energy barrier $\Delta E \propto \kB T a^2/
\<u^2\>$ to be overcome, one recovers the 
Hall--Wolynes equation 
\[
\tau \propto \exp(a^2/2 \<u^2\>);
\]
this connects the structural relaxation time
to a short--time elastic property, the DW factor. 
This approach has recently been~\cite{Larini2008} generalized by introducing a
probability distribution for $a^2$, and successfully tested
against experimental and numerical data, including both strong and fragile 
glass--formers. The main message from this work is that there is a
one--to--one correspondence between the relaxation dynamics and the DW
factor.

These two approaches for understanding glassy relaxation times that we
have described above are close related to each other because the DW factor
is fixed mainly by the shear modulus of the
material~\cite{DyreOlsen2004}. We will therefore refer to them
collectively as ``elastic models''.

In this paper we consider the applicability of elastic models to
suspensions of particles interacting via a harmonic potential.
Similar finite range purely repulsive potentials are of interest as model
for the interaction of macroscopic particles such as
bubbles, foams and microgels, whose dynamics exhibit glassy features
at high concentration and/or low temperature. 
In addition, these potentials are also of interest for being able to
give rise to water--like density anomalies 
at high
densities~\cite{Frenkel2009,Berthier2010,Wang2012,CiamarraSollich_SM,CiamarraSollich_JCP},
in spite of their manifest simplicity.
The possible applicability of elastic models to these systems is of
particular interest because elastic models are based on an energy landscape interpretation
of the dynamics, while density anomalies have been rationalized
by entropic arguments~\cite{Berthier2010}.

We will show that in finite range repulsive systems the shear modulus
estimated from the properties of the underlying energy landscape 
overestimates the plateau shear modulus, which implies
that the energy landscape properties are poorly correlated with the elasticity.
Nevertheless, we do find correlations between relaxation time,
diffusivity,
plateau shear modulus and DW factor that are consistent with those predicted by the elastic
model of the glass transition. Our results indicates that while elastic models
correctly describe the relaxation dynamics of the system,
their interpretation in terms of features of the energy landscape 
should be reconsidered.

\section{Model}
We consider a polydisperse mixture of $N = 10^3$ harmonic disks of mass $m$, in two dimensions. 
Diameters are uniformly distributed in the range $[D_{\rm min}:D_{\rm max}]$, with
the difference $D_{\rm max}-D_{\rm  min}$ between the largest and
smallest diameter being 82\% of the mean diameter $(D_{\rm max}+D_{\rm
  min})/2$ so that the distribution is fairly broad; this is necessary to
prevent crystallisation at high volume fractions.
Two particles $i$ and $j$ with average diameter $D=(D_i+D_j)/2$ and at a distance $r$
interact via a potential 
\begin{equation}
 v(r) = 
 \begin{cases}
  \frac{1}{2} \epsilon \left(\frac{D-r}{D_{\rm max}}\right)^2 &\mbox{if } r \leq D \\
  0 & \mbox{if } r > D
 \end{cases}
 \label{eq:potential}
\end{equation}
In the following, lengths, masses and energies are expressed in units of $D_{\rm max}$, $m$ and of $\epsilon$, respectively,
and the density is expressed via the volume (or rather, area) fraction
$\phi = N \<A\>/L^2$. Here
$L$ is the system size, $\<A\>$ the average particle area, and $N$ the
number of particles. 
We have performed molecular dynamics simulations~\cite{lammps} at fixed volume,
temperature and particle number, integrating the equations of motion using
the Verlet algorithm, and constraining the temperature via a Nose--Hoover thermostat.
The shear stress is defined as the off-diagonal term of the stress tensor,
\begin{equation}
\sigma_{xy} = \frac{1}{V} \left(\sum_{i=1}^N m v_{xi}v_{yi} + \frac{1}{2}\sum_{i \neq j} r_{xij}F_{yij} \right),
\label{eq:sxy}
\end{equation}
where $v_{\alpha i}$,  $F_{\alpha ij}$ and $r_{\alpha ij}$ are the $\alpha$-components of the velocity of the $i$-th particle,
of the force between particles $i$ and $j$ and of the separation
between them. The transient shear modulus is related
to the decay of the shear stress fluctuations, $G(t) = \frac{V}{\kB T} \< \sigma_{xy}(0) \sigma_{xy}(t) \>$. 
We explore the features of the underlying energy landscape by repeatedly minimizing the energy 
of the system via the conjugate--gradient protocol to find the
instantaneous inherent structure, and measuring the 
inherent shear stress $\sigma_{xy}^{\rm IS}$. The latter is computed via Eq.~\ref{eq:sxy},
where all velocities are set to zero. The transient inherent structure shear modulus is
defined as  $G^{\rm IS}(t) = \frac{V}{k_B T} \< \sigma^{\rm IS}_{xy}(0) \sigma^{\rm IS}_{xy}(t) \>$, where 
$T$ is the temperature of the parent liquid; $G^{\rm IS}(0)$ is then
the same as $G^{\rm IS}$ defined
above.

\section{Density anomalies}
\begin{figure}[!t]
\begin{center}
\includegraphics*[scale=0.33]{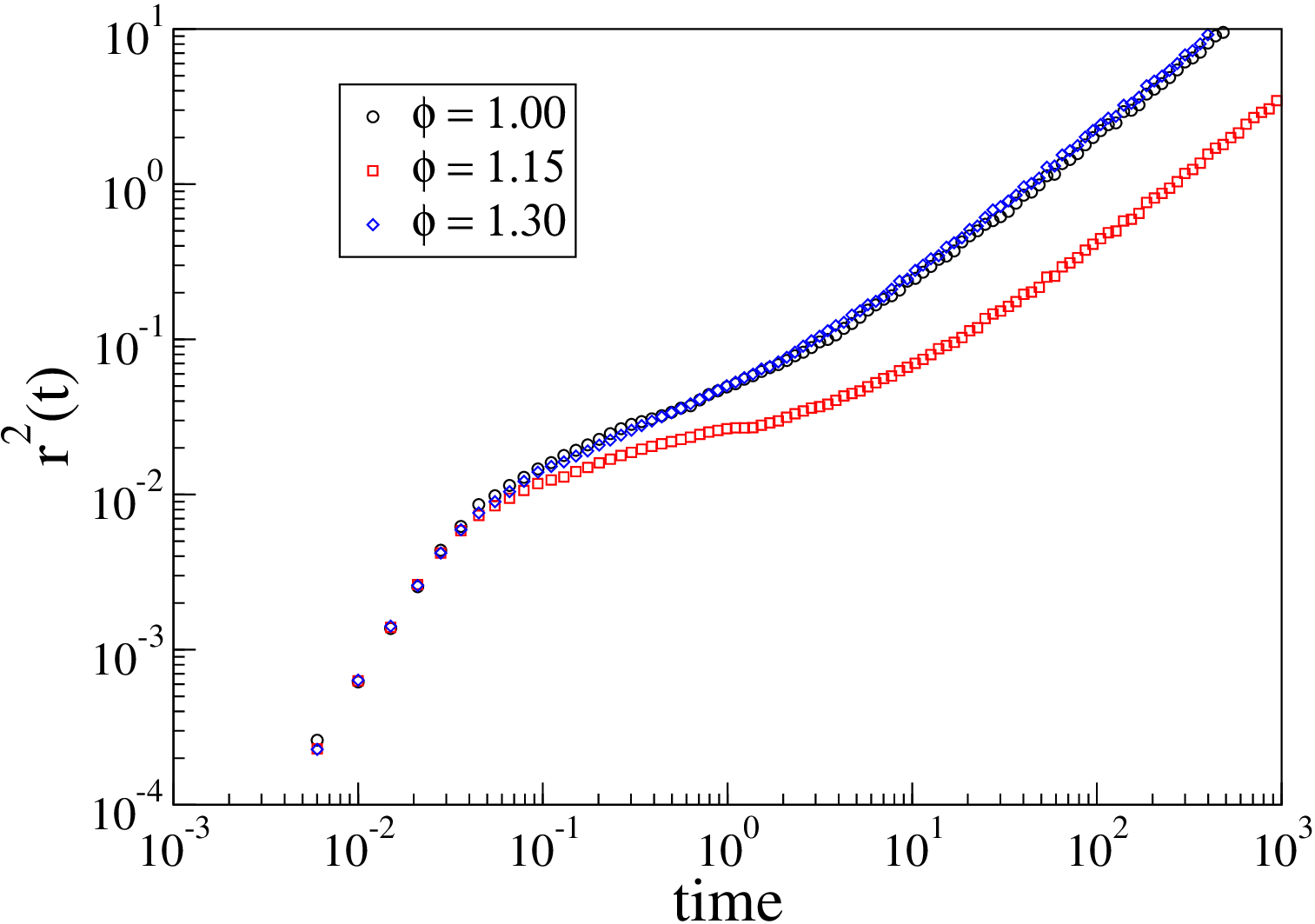}
\end{center}
\caption{\label{fig:msd}
(color online) Mean square displacement for a system of $N = 1000$ particles
at $T = 0.18$ and $\phi = 1$, $1.15$ and $1.3$.
}
\end{figure}
\begin{figure}[!h]
\begin{center}
\includegraphics*[scale=0.30]{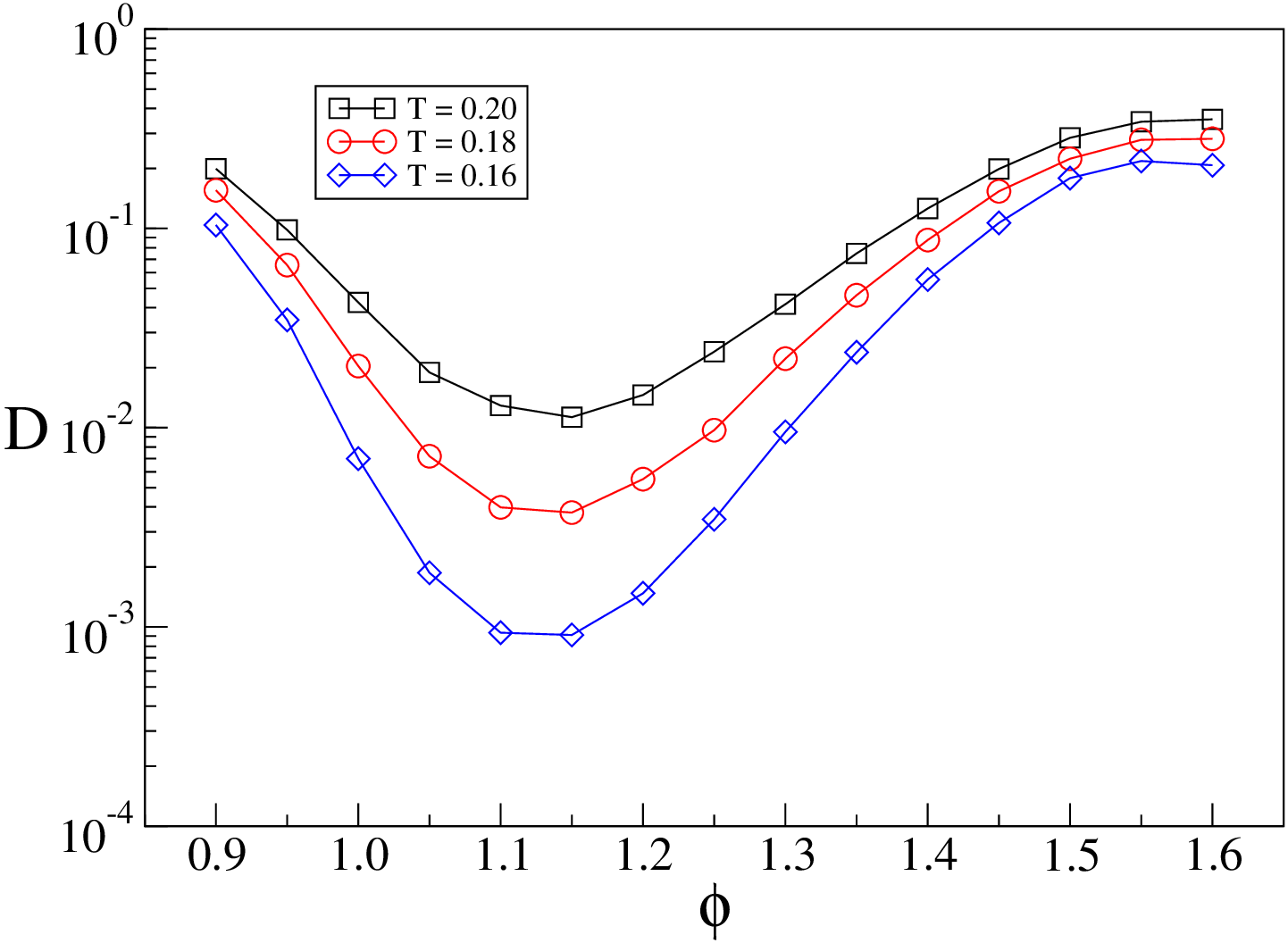}
\end{center}
\caption{\label{fig:diffanomaly}
Volume fraction dependence of the isothermal diffusivity, for different
values of the temperature as indicated.
}
\end{figure}

Systems of harmonic and Hertzian particles are 
characterized by water--like density anomalies~\cite{Frenkel2009,Berthier2010,Wang2012,CiamarraSollich_SM,CiamarraSollich_JCP},
including a non--monotonic variation of the diffusivity upon isothermal compression,
as well as a negative thermal expansion coefficient. 
As an example, we show in Fig.~\ref{fig:msd} the mean square displacement 
for three different values of the volume fraction, $\phi = 1$, $\phi = 1.15$ and $\phi = 1.3$,
at $T = 0.18$. 
Fig.~\ref{fig:diffanomaly} shows how the corresponding diffusivities $D$
depend on volume fraction and temperature. One sees that for
$T=0.18$, $\phi=1.15$ gives essentially the lowest diffusivity;
for $\phi=1$ one has standard behaviour, with $D$ decreasing as
density increases, while for $\phi=1.3$ the trend is reversed and one
has anomalous behaviour.
Fig.~\ref{fig:msd} suggests that the dynamics
at $\phi = 1$ and at $\phi = 1.3$ are very similar, with the two mean
square displacement curves visually indistinguishable; we return
to this point later on.

It is of particular interest to investigate whereas elastic models of the glass transition capture
the observed density anomalies. Indeed, density anomalies are mainly driven by
the density dependence of the entropy of the system, while elastic models are based
on a purely energetic interpretation of the dynamics.

\section{Inherent and liquid shear elasticity}
\begin{figure}[t!]
\begin{center}
\includegraphics*[scale=0.30]{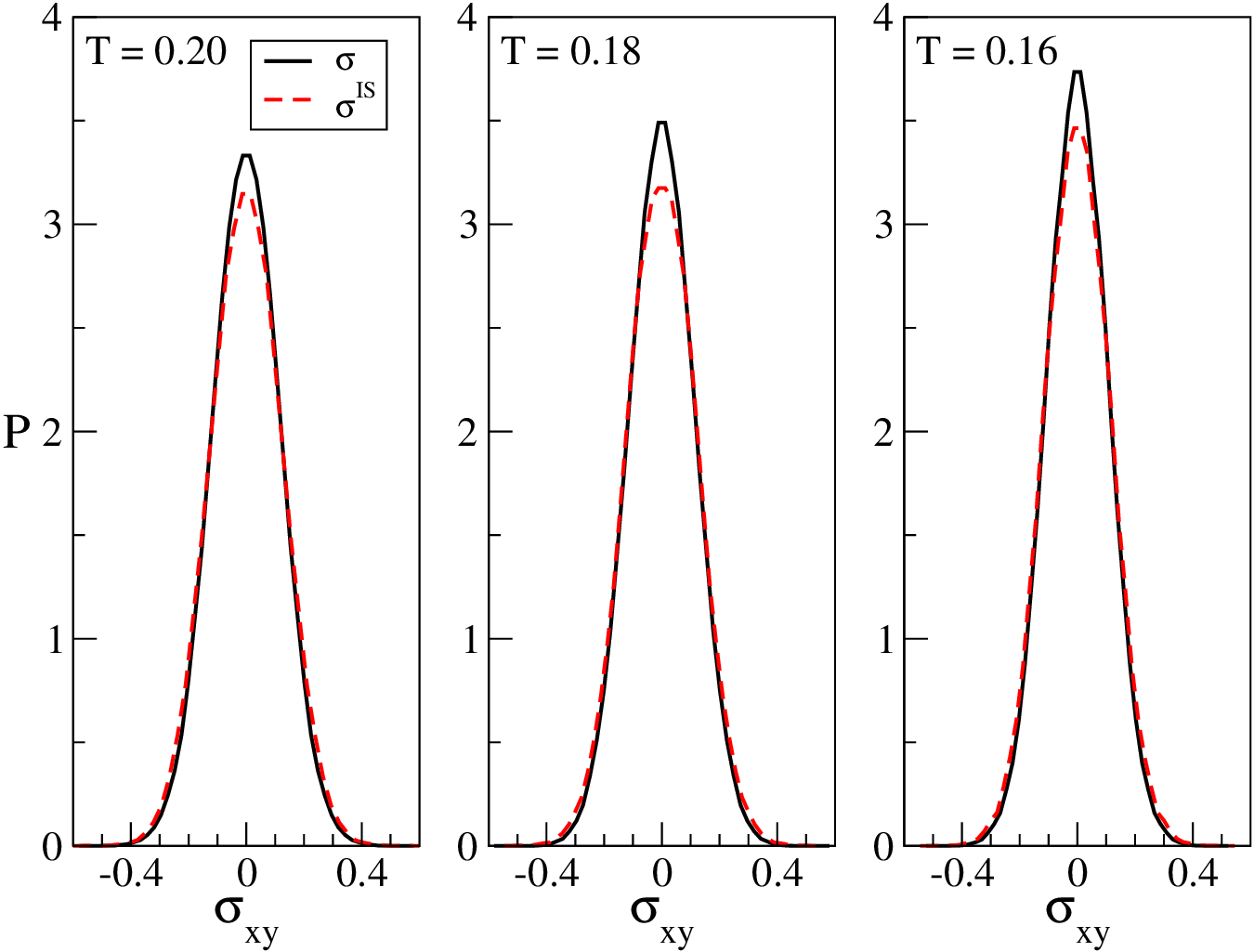}
\end{center}
\caption{\label{fig:PsxyT}
(color online) Probability distribution of the shear stress and of the inherent shear stress,
at $\phi = 1.15$, for different temperatures as indicated.
}
\end{figure}

\begin{figure}[t!]
\begin{center}
\includegraphics*[scale=0.30]{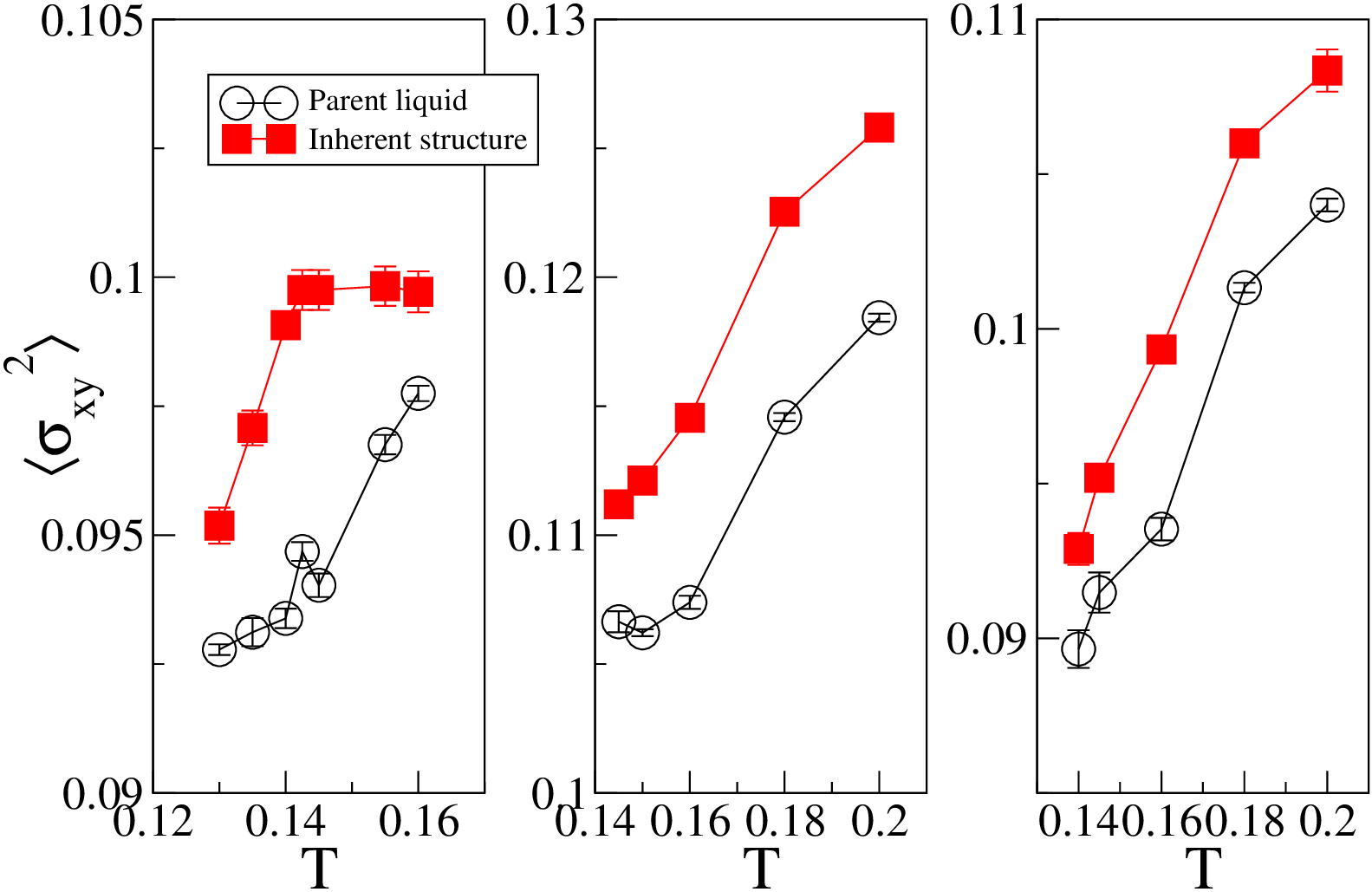}
\end{center}
\caption{\label{fig:GvsT}
(color online) Temperature dependence of the variance of the shear stress for the 
parent liquid and the inherent structures, for $\phi = 1$ (left panel),
$\phi = 1.15$ (central panel) and $\phi = 1.3$ (right panel).
}
\end{figure}

Elastic models~\cite{Dyre06,DyreOlsen2004} of the glass transition are 
predicated on the assumption that the dynamics of supercooled liquids
consists of a series of
jumps between local potential energy minima, i.e.\ inherent structures.
The local curvature of these minima, which is related to the shear modulus of the system,
sets the energy scale of these events and thus controls the relaxation dynamics.
This energy--landscape interpretation makes it of interest to investigate the relation between the elastic
properties measured in the liquid phase, and those observed after quenching the system to its inherent structure.
We have therefore measured the probability distribution function of the shear
stress $\sigma_{xy}$ and of the inherent shear stress, $\sigma_{xy}^{\rm IS}$, 
whose variances are proportional to the instantaneous shear moduli.
Fig.~\ref{fig:PsxyT} shows that these distributions have a
Gaussian--like shape at all temperatures. 
In systems interacting via inverse power--law
potentials~\cite{Abraham} $\< \sigma_{xy}^2 \>$
decreases on cooling, and approaches the value of $\< (\sigma^{\rm IS}_{xy})^2 \>$
which is temperature independent.
We find that also in our system, $\< \sigma_{xy}^2 \>$ decreases on cooling.
However, as is clear from Fig.~\ref{fig:GvsT}, 
significant differences appear in the behaviour of the inherent structure stresses, 
and the relative size of the instantaneous and inherent structure stresses.

We note first that $\< (\sigma^{\rm IS}_{xy})^2 \>$ depends on the temperature
of the parent liquid. This behaviour signals the fact that
the system explores different regions of the energy landscape at different temperatures.
In similar systems, this feature has recently been exploited to prove that
the volume fraction at which the inherent structure shear stress vanishes,
the jamming volume fraction, depends on the temperature of the parent liquid~\cite{Chaudhuri}.
This proves that the jamming volume fraction is protocol dependent~\cite{Chaudhuri,Ciamarra2010}.
At high temperatures the dynamics of the system is no longer affected 
by the energy landscape, and $\< (\sigma^{\rm IS}_{xy})^2 \>$ should be 
temperature independent. Fig.~\ref{fig:GvsT} suggests that we have reached
this high temperature regime for $\phi = 1$.

\begin{figure}[t!]
\begin{center}
\includegraphics*[scale=0.30]{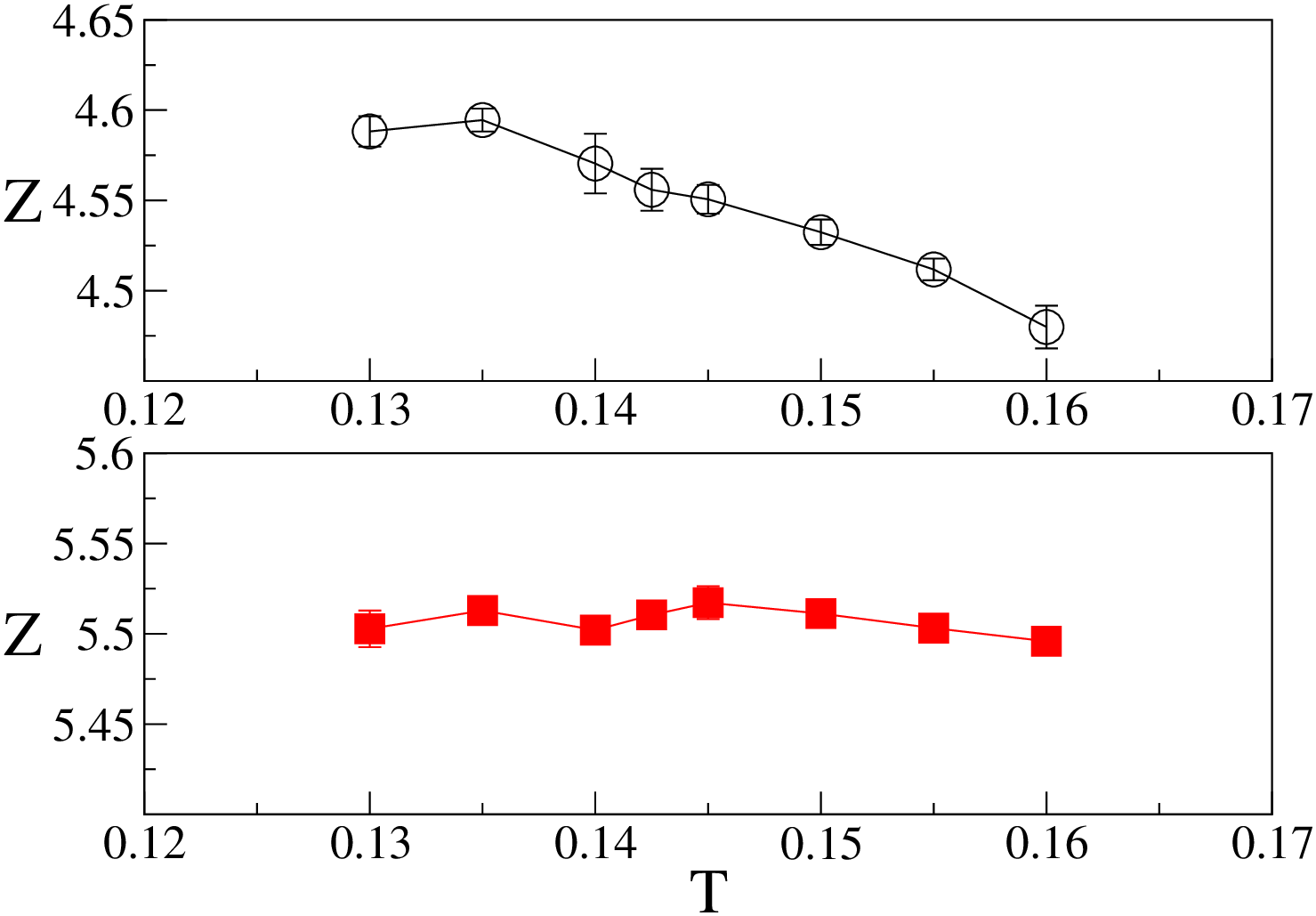}
\end{center}
\caption{\label{fig:ZT}
(color online) Temperature dependence of the average contact number per particle
for the parent liquid (top) and the inherent structures (bottom), for $\phi = 1.00$.
}
\end{figure}

Looking next at the relation between instantaneous and inherent structure stresses, we find consistently (see Fig.~\ref{fig:GvsT}) 
that $\< (\sigma^{\rm IS}_{xy})^2 \> > \< \sigma_{xy}^2 \>$,
while the opposite relation is found in inverse--power law liquids.
We can make sense of this result by considering that the fluctuations of the shear stress
are related to the instantaneous shear modulus. For harmonic potentials,
whose second (radial) derivative of the potential is constant, the instantaneous shear
modulus is expected to scale with the density of contacts, $G \propto z \phi$ where $z$ is the
average contact number per particle. Fig.~\ref{fig:ZT} shows the
temperature dependence of $z$ for the parent liquid, where $z$ increases
on cooling, and for the inherent structures, where $z \approx 5.5$ is constant.
At all temperatures, the average contact number of the inherent structures
is larger than the average contact number of the parent liquid.
This implies that the inherent shear modulus
$G^{\rm IS} = \frac{V}{k_B T} \< (\sigma^{\rm IS})^2 \>$,
is larger than the shear 
shear modulus of the parent liquid, $G(0) = \frac{V}{k_B T} \< \sigma_{xy}^2 \>$.
Accordingly, $\< (\sigma^{\rm IS})^2 \> > \< \sigma_{xy}^2 \>$, which is the relation we saw in Fig.~\ref{fig:GvsT}. 
In power law potential systems the situation is different:
the single particle bulk modulus, which is directly related to the second derivative of the potential, 
is not constant but increases without bound
as the interparticle distance becomes smaller.
Thus, if the average interparticle distance {\em increases} due to more efficient packing when the system is quenched to its the inherent
structure, then the shear modulus of the inherent structures is expected to be smaller
than that of the parent liquid.

\section{Elastic models}
\begin{figure}[t!]
\begin{center}
\includegraphics*[scale=0.30]{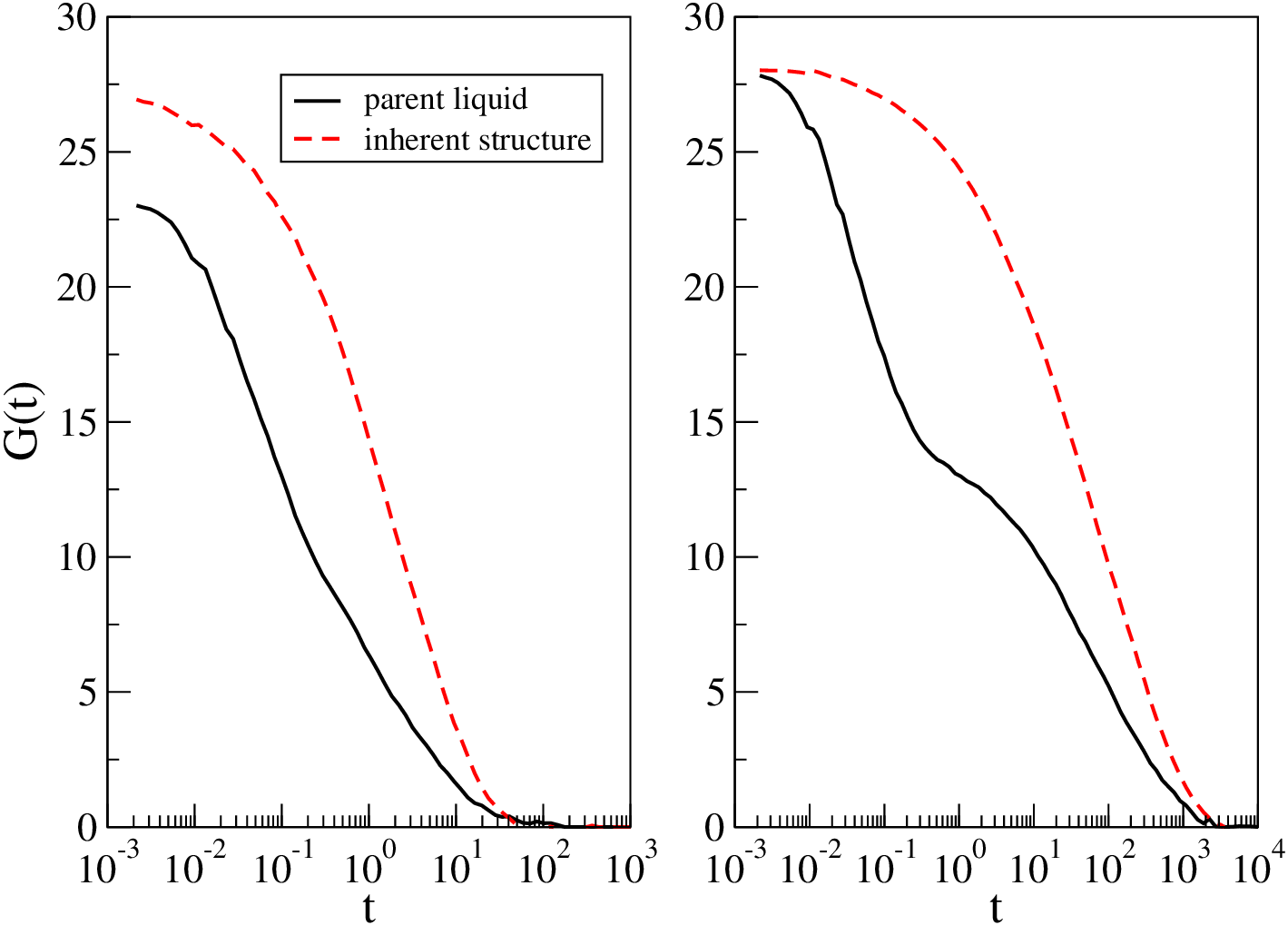}
\end{center}
\caption{\label{fig:Gt}
(color online) Transient shear modulus and transient inherent structure shear modulus,
for $\phi = 1$ and $T = 0.16$ (left panel) and $T = 0.13$ (right panel).
}
\end{figure}

We now consider whereas the slowdown of the dynamics is well described
by elastic models of the glass transition. First we consider
Dyre's shoving model, according to which 
$\log \tau \propto G_pV_{\rm at}/\kB T$, where as before
$G_p$ is the plateau shear modulus, and $V_{\rm at}$
a local activation volume. As usual we assume that the latter does not change with temperature, 
but now add the assumption that it is also independent of volume fraction.
Fig.~\ref{fig:Gt} shows the relaxation
dynamics of the instantaneous shear stress: the transient shear modulus $G(t)$ exhibits the two--step decay typical of glasses. 
From $G(t)$, we can estimate a plateau modulus $G_p$ in the deeply supercooled regime
as the value of $G(t)$ 
at the (intermediate) time at which the derivative of $G(t)$ with respect to 
$\log(t)$ is minimal. The stress relaxation time $\tau$ can then be extracted via a stretched exponential fit of the final decay of $G(t)$.

\begin{figure}[t!]
\begin{center}
\includegraphics*[scale=0.30]{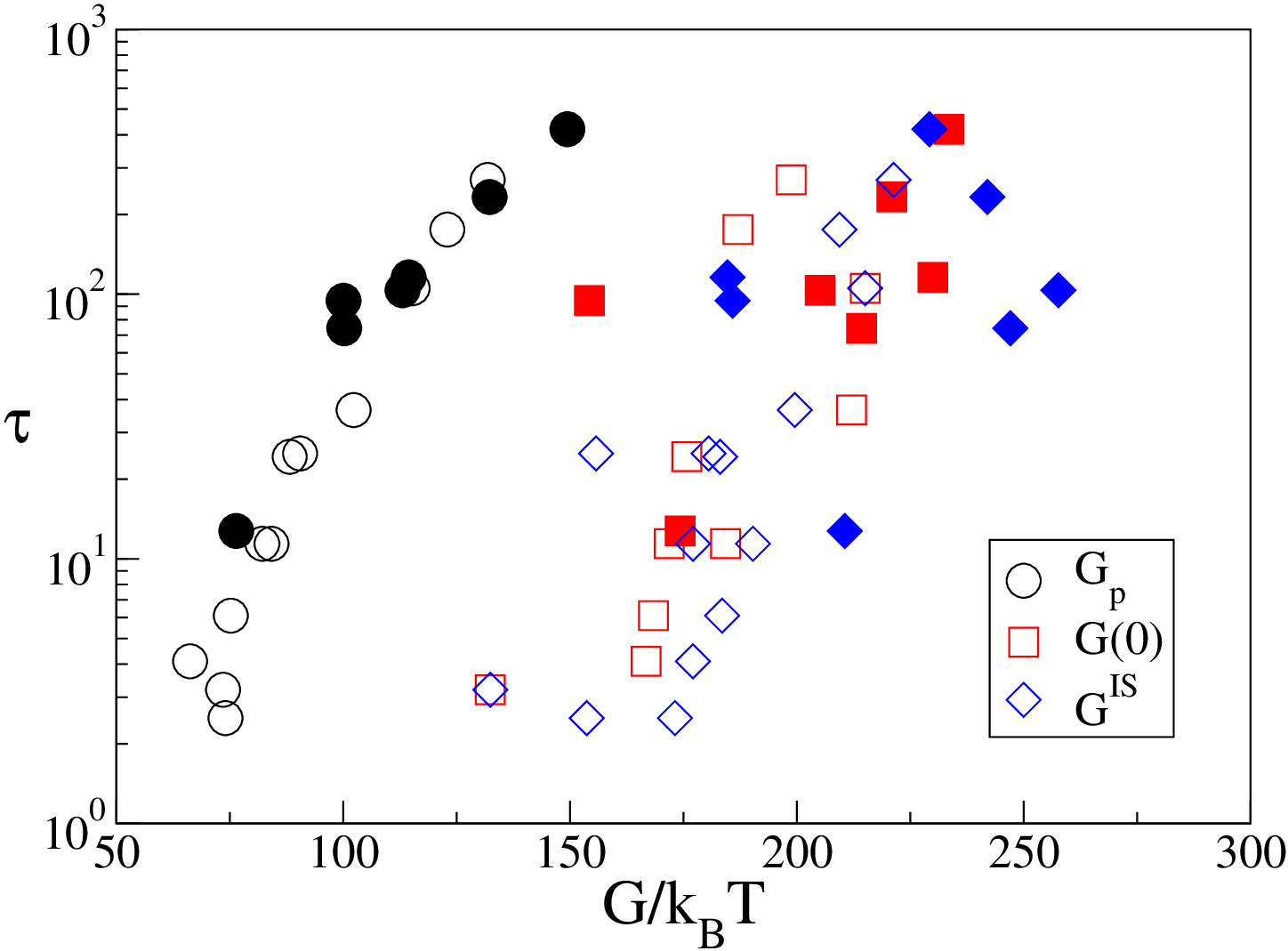}
\end{center}
\caption{\label{fig:tauG}
(color online) Correlation of the relaxation time $\tau$ with the 
plateau shear modulus $G_p$, the instantaneous shear modulus $G(0)$, and instantaneous inherent structure shear modulus $G^{\rm IS}$.
Data for a range of different temperatures and volume fractions are shown.
The filled symbols refer to $T = 0.14$ and volume fractions $\phi$ in the region of the density anomaly. 
}
\end{figure}

Having obtained data for $G_p$ and $\tau$, we can then assess the applicability of Dyre's shoving model to our system: see Fig.~\ref{fig:tauG}.
Here we show (black circles) that the relaxation time is correlated fairly tightly with the plateau shear modulus $G_p$ 
In particular, the expected exponential relation between relaxation time and plateau
modulus divided by $\kB T$ is observed in the deeply supercooled regime.
It is important to note here that the figure combines points taken at different temperatures
and volume fractions. In particular, the filled black circles refer to 
$T = 0.14$ and volume fractions around the one giving the  minimal diffusivity. This
proves that elasticity as measured by the plateau shear modulus is well correlated with the dynamics
in the anomalous region.

While the prediction of the shoving model appears to be reasonably well verified,
its interpretation in terms of the features of the energy landscape of the system is not.
Indeed, in this interpretation the plateau shear modulus should be related
to the inherent structure shear modulus, as recently found in
power--law liquids~\cite{Puosi2012,Abraham}. In these systems  $G^{\rm IS}(0) \approx G_p$ and at all times 
$G^{\rm IS}(t) \leq G(t)$
Intuitively, the instantaneous stress has larger fluctuations over short time scales than the inherent structure stress, 
but around the timescale of the plateau the fast fluctuations have averaged out and the relaxations of 
instantaneous and inherent structure stress track each other.

In our short range harmonic repulsive systems, on the other hand, we find $G^{\rm IS}(0) > G(0)$ 
as shown in Fig.~\ref{fig:Gt} and 
as explained in the previous section. Since $G(0)>G_p$, this implies also $G^{\rm IS}(0) > G_p$: 
the approximate equality between these quantities no longer holds. 
In fact we find that $G^{\rm IS}(0)$ is not even proportional to $G_p$. 
This is clear indirectly from Fig.~\ref{fig:tauG}, as no data collapse is found when $\tau$ is plotted
versus $G^{\rm IS}(0)/\kB T$ instead of $G_p/\kB T$; the same holds if we use $G(0)$ instead of $G^{\rm IS}(0)$. 
This result clarifies that,
while the dynamics of harmonic systems are determined by their elastic properties,
these properties are not related in a simple way to those of the energy landscape.
This is consistent with a previous entropic interpretation
of the observed density anomalies~\cite{Berthier2010}.

We next consider, as a test of the second elastic model mentioned in the introduction, whereas the relaxation dynamics is also closely
correlated with the DW factor.
Fig.~\ref{fig:msd} suggests that this may well be the case, as
for $T = 0.18$ we saw that the mean square displacements
at $\phi = 1$ and at $\phi = 1.3$ coincide to good accuracy at all times, consistent the macroscopic diffusivities
also being equal between these two volume fractions.
We have therefore investigated the existence of correlations between the diffusion coefficient $D$
and the DW factor, $\db$. 
Operatively~\cite{Larini2008}, we define $\db = \<r^2(t_{DW})\>$, where $t_{DW}$
is the time of minimal diffusivity. We determine this time by considering the time dependence of the diffusivity 
exponent $b(t) = \partial \log(\<r^2(t)\>)/\partial \log(t)$,
which varies in time from the value $b =2$, characteristic
of the short time ballistic motion, to the value $b = 1$ for the long
time diffusive motion. In the supercooled regime, the mean square displacement at intermediate
times is sub-diffusive, and $b(t) < 1$ has a minimum at some time $t_{DW}$.

\begin{figure}[t!]
\begin{center}
\includegraphics*[scale=0.30]{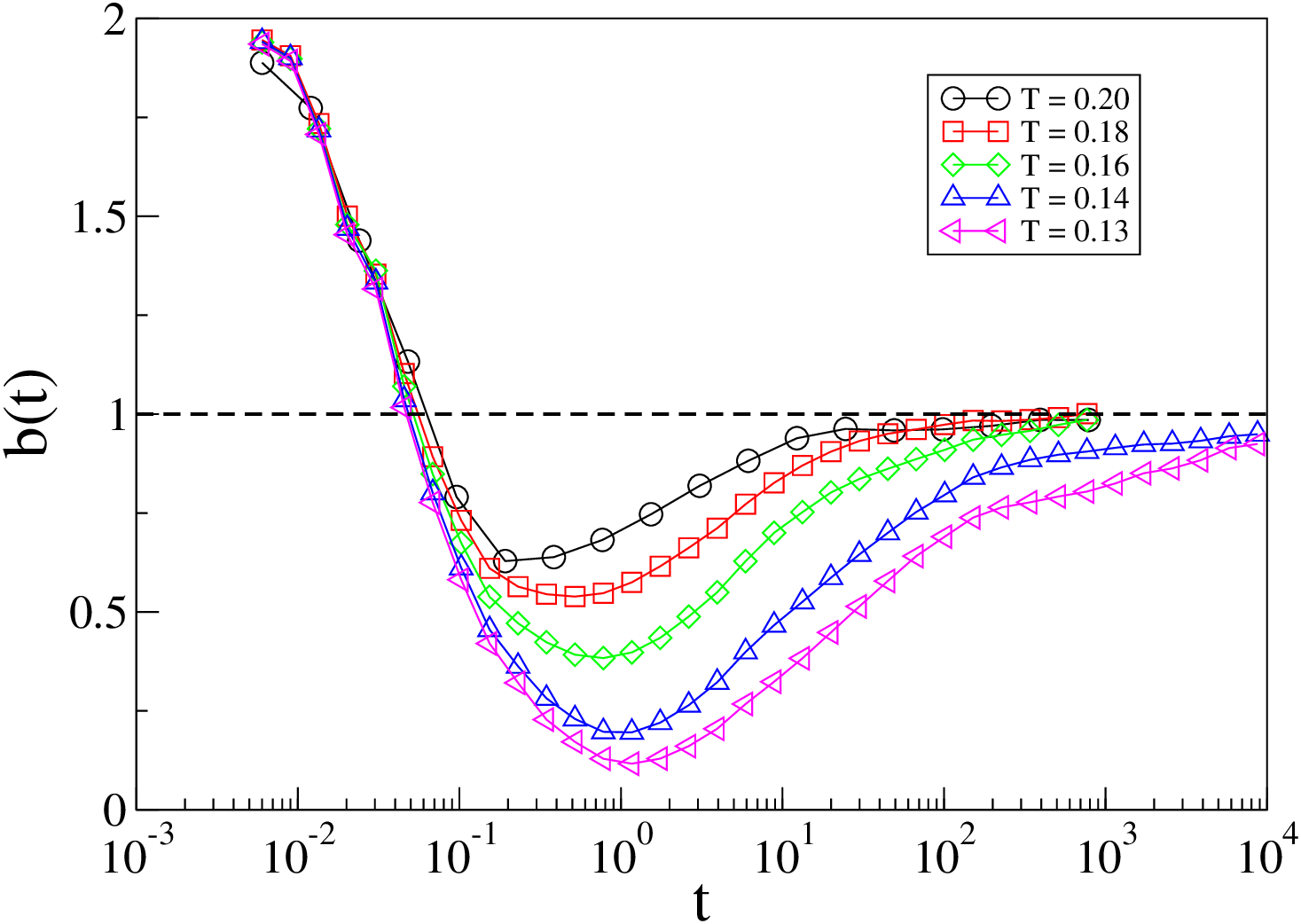}
\end{center}
\caption{\label{fig:bt}
(color online) Time evolution of the exponent $b(t)$ characterizing the diffusivity of the system,
$\<r^2(t)\> \propto t^{b(t)}$, at different temperatures for volume fraction $\phi= 1$.
}
\end{figure}

\begin{figure}[t!]
\begin{center}
\includegraphics*[scale=0.30]{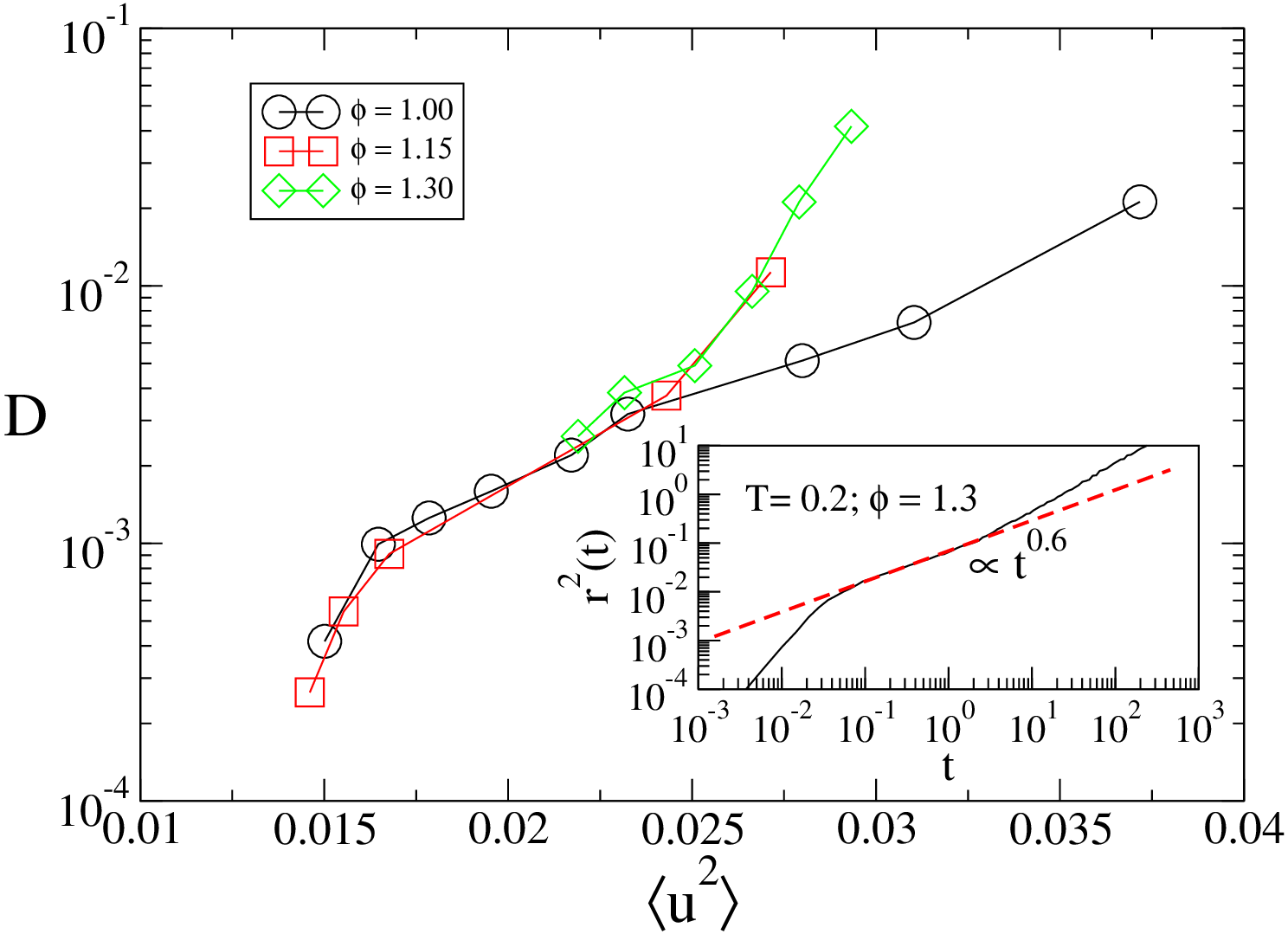}
\end{center}
\caption{\label{fig:DU2}
(color online) Parametric plot of the diffusion coefficient versus the Debye-Waller factor,
for $\phi = 1, 1.15$ and $1.3$. In the deeply supercooled regime the diffusivity appears
to be determined by $\db$. At higher temperatures the determination
of $\db$ is affected by substantial uncertainty because the subdiffusive regime disappears; in
addition, at high volume fraction $\<r^2(t)\>$ can develop an extended subdiffusive regime 
as illustrated in the inset.
}
\end{figure}

Fig.~\ref{fig:DU2} displays the resulting dependence of the diffusion coefficient $D$ on $\db$.
In the deeply supercooled regime of low $D$, the figure suggests that $D$ is uniquely determined by $\db$.
At higher temperatures this is no longer the case. We note, however,
that at higher temperatures the identification of $\db$ is subject to large errors
as the subdiffusive regime disappears. In addition,
in the anomalous volume fraction range we observe the presence of a long subdiffusive regime 
with a nearly constant subdiffusive exponent, as illustrated in the inset of Fig.~\ref{fig:DU2}.

\begin{figure}[!t]
\begin{center}
\includegraphics*[scale=0.33]{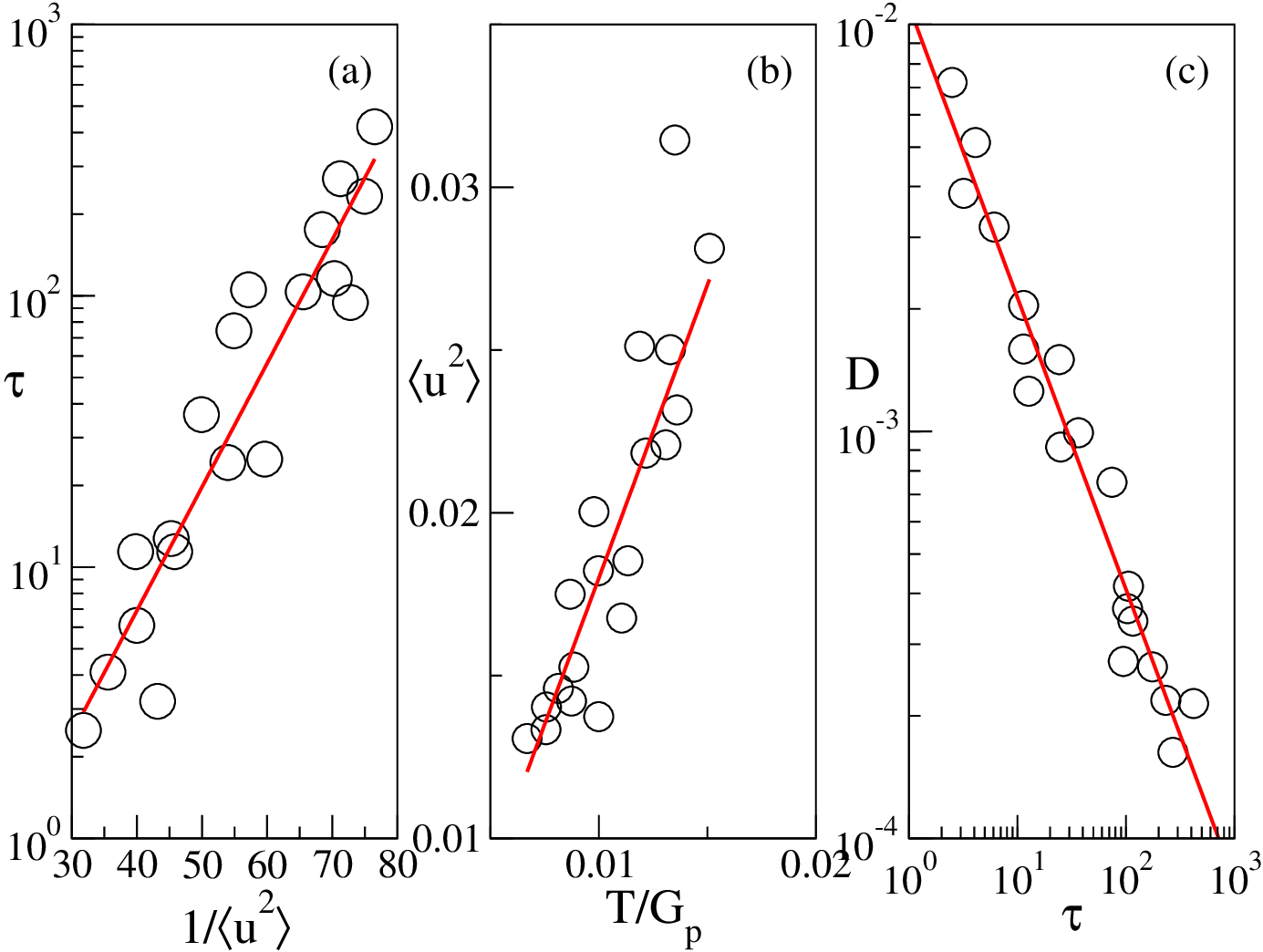}
\end{center}
\caption{\label{fig:comparison}a
(color online) Correlations between (a) relaxation time and DW factor, (b) DW factor
and plateau modulus, and (c) diffusivity and relaxation time. 
The straight line in panel (c) is $D \propto \tau^{-q}$, with $q \approx 0.72$. 
Data points shown cover a range of different temperatures and volume fractions.
}
\end{figure}

We conclude this section by returning to the point that
the two elastic models we have considered are not independent because $\db$
is (mainly) determined by the shear elasticity~\cite{DyreOlsen2004}.
Indeed, we do also find a clear correlation between the relaxation
time and the DW factor, which is consistent
with the Hall--Wolynes equation, $\tau \propto \exp(a^2/2 \<u^2\>)$
as shown in Fig.~\ref{fig:comparison}a.
The combined validity of this relation and of Dyre's model
implies the relation $T/G_p \propto \db$, which is also compatible
with our numerical data as Fig.~\ref{fig:comparison}b shows.
Finally, we note (Fig.~\ref{fig:comparison}) that we also find a relation
between diffusivity and relaxation time, 
with $D \propto \tau^{-q}$ and $q \approx 0.72$.

\section{Conclusions}
We have demonstrated that elastic models of the glass transition correctly capture the
slow dynamics of harmonic particle systems, in the temperature and
volume fraction region where density anomalies are observed. 
However, we have found the relevant elastic constants
not to be related to the features of the energy landscape of the system.
This result challenges the usual potential energy landscape interpretation of elastic models,
and suggests that they should generally be interpreted by referring to the 
free energy landscape of the system. Future directions include 
an investigation of the validity of elastic models in other liquids
with density driven anomalies, such as the Gaussian potential, the Jagla potential,
or water--like model, as well as in liquids with temperature driven anomalies
such as the sticky hard--sphere model.

\end{document}